\documentclass[12pt]{article}

\def\map{\mbox{\boldmath $\phi$}}
\def\be{\begin{equation}}
\def\ee{\end{equation}}
\def\la{\langle}
\def\ra{\rangle}
\def\IP{\hbox{\rm I\kern -1.6pt{\rm P}}}
\def\IC{{\hbox{\rm C\kern-.58em{\raise.53ex\hbox{$\scriptscriptstyle|$}}
    \kern-.55em{\raise.53ex\hbox{$\scriptscriptstyle|$}} }}}
\def\IN{\hbox{I\kern-.2em\hbox{N}}}
\def\IR{\hbox{\rm I\kern-.2em\hbox{\rm R}}}
\def\ZZ{\hbox{{\rm Z}\kern-.3em{\rm Z}}}
\def\IT{\hbox{\rm T\kern-.38em{\raise.415ex\hbox{$\scriptstyle|$}} }}
\def\notsub{\hbox{$\subset$\kern-.55em\hbox{/}}}
\def\i{{\rm i}}
\newcommand{\f}[1]{{f_{#1}^{i_{#1}}}}

\newtheorem{theorem}{Theorem}
\newtheorem{lemma}[theorem]{Lemma}

\newtheorem{conjecture}[theorem]{Conjecture}

\begin{document}
\title{The Burnett expansion of the periodic Lorentz gas}
\author{
C. P. Dettmann\thanks{Department of Mathematics, University of Bristol,
University Walk Bristol BS8 1TW, UK}}

\maketitle

\begin{abstract}
Recently, stretched exponential decay of multiple correlations in the
periodic Lorentz gas has been used to show the convergence of a series
of correlations which has the physical interpretations as the fourth
order Burnett coefficient, a generalization of the diffusion coefficient.
Here the result is extended to include all higher order Burnett coefficients,
and give a plausible argument that the expansion constructed from the Burnett
coefficients has a finite radius of convergence.
\end{abstract}

\section{Introduction}
The {\em Lorentz gas} is a model used in statistical mechanics, consisting of a
point particle moving at constant velocity except for specular collisions
with smooth (specifically $C^3$) convex fixed scatterers in $d\geq 2$
dimensions.  The original model~\cite{L} has randomly placed scatterers
in infinite space, and is thought to have power law decay of correlations,
so that the Burnett coefficients (defined below as sums of such correlations)
are not generally expected to exist~\cite{DC,vB}. Here we consider a periodic
arrangement of scatterers which is equivalent to a dispersing billiard on
a torus, for which it is known that two-time correlations of the discrete
(collision) dynamics decay exponentially~\cite{C99,Y}.
This, together with the {\em finite horizon} condition, that is, that
the time between collisions is bounded, implies the existence of the
diffusion coefficient ($D^{(2)}$ below).

A recent paper gives a stretched exponential decay
of multiple correlations~\cite{CD}, and uses this to show
(again with finite horizon) that the fourth order Burnett
coefficient ($D^{(4)}$ below) exists.  Here we extend this result to
all the Burnett coefficients.  A common example for $d=2$ with a finite horizon
is given by circular scatterers on a triangular lattice; for $d>2$ the
finite horizon condition requires either nonspherical scatterers, or more
than one scatterer per unit cell. The Lorentz gas and a number of extensions
are discussed in Ref.~\cite{S}.

The Burnett coefficients $D^{(m)}$ discussed in
this paper are defined using series of correlation functions.  Section 2
defines these series and gives three basic results about them.  Section
3 gives the main result of this paper, the proof of convergence of these
series.  The series arise in a physical description of diffusion, however the
derivation involves hydrodynamic approximations and interchange of
limits which have not been justified rigorously for the Lorentz gas.
The physical motivation together with the non-rigorous derivation of the
series given below from previously stated formulas is given in the
final section, together with a conjecture about the Burnett expansion.

The author is grateful for helpful discussion with N. I. Chernov, E. G. D.
Cohen, J. R. Dorfman and P. Gaspard, and for the support of the Engineering
Research Program of the Office of Basic Energy Sciences at the US Department
of Energy, contract \#DE-FG02-88-ER13847, and the Nuffield Foundation,
grant NAL/00353/G.

\section{Definitions}
In the following, $\map(x)$ is the billiard map defined on the collision space
$M$, consisting of points $x=({\bf r},{\bf v})\in M$ for which the position
${\bf r}\in \IR^d$ is on the boundary of one of the scatterers and the
velocity following a collision ${\bf v}\in\IR^d$ is of unit magnitude in
an outward direction from the
scatterer.  Greek indices $\alpha,\beta,\ldots=1,\ldots,d$ denote components
of vectors and tensors in $\IR^d$, and a dot ${\bf a}\cdot{\bf b}$ denotes
the usual inner product $\sum_\alpha a_\alpha b_\alpha$ corresponding to the
Euclidean metric.  We have two
functions $T:M\to \IR$ and ${\bf a}:M\to \IR^d$ which describe the embedding
of the collision dynamics into physical space and time, as follows.
$T(x)$ is the time (also distance since the speed of the
particle is one) between the collision at $x$ and the next; it is a piecewise
H\"older continuous function~\cite{BSC,C94}. ${\bf a}(x)$ is the lattice
translation vector associated with this free flight when the configuration
variable $\bf r$ is unfolded onto a periodic tiling of $\IR^d$; it is
a linear combination of the lattice basis vectors ${\bf e}^{(\alpha)}$
with integer coefficients, and is a piecewise constant function.  
The finite horizon condition ensures that both $T$ and ${\bf a}$ are bounded.
The average $\la\cdot\ra$ denotes integration over $M$ with
respect to the invariant equilibrium measure.  In terms of this average we
define $\Delta T:M\to \IR$ by $\Delta T(x)=T(x)-\la T \ra$ so that
$\la \Delta T \ra=0$.

The billiard dynamics is time reversal invariant, that is, there exists an
involution ${\cal T}:M\to M$ (given simply by the specular reflection law)
with the property
\begin{equation}
\map\circ{\cal T}\circ\map={\cal T}
\end{equation}
In addition, $\cal T$ preserves the equilibrium measure, that is,
\begin{equation}
\langle {\cal T}\circ g\rangle=\langle g\rangle
\end{equation}
for arbitrary measurable function $g:M\to M$.
The map $\cal T$ also satisfies
\begin{eqnarray}
T\circ{\cal T}\circ\map&=&T\\
{\bf a}\circ{\cal T}\circ\map&=&-{\bf a}
\end{eqnarray}
Thus $\la {\bf a} \ra=0$.
 
The wave vector $\bf k$ is to
be understood as a formal real expansion parameter with $d$ components
(although physically we would like to interpret it as a vector with a
value in $\IR^d$).  The {\em dispersion relation} $s[{\bf k}]$ is to be
understood as a formal power series
\be s[{\bf k}]=\sum_{m=2}^{\infty}\i^m\sum_{\alpha_1\ldots\alpha_m}
D^{(m)}_{\alpha_1\ldots\alpha_m}k_{\alpha_1}\ldots k_{\alpha_m}
\label{e:disp} \ee
in terms of the {\em Burnett coefficients} $D^{(m)}$ which are assumed to be
real, totally symmetric tensors of rank $m$.  That is, an equation
(specifically Eq.~(\ref{e:Bps}) below) involving
$s[{\bf k}]$ is to be interpreted as a sequence of equations (specifically
Eq.~(\ref{e:Bst}) below) obtained by
equating coefficients of powers of $\bf k$.  The symbol $\i$ denotes
$\sqrt{-1}$.

The existence of Burnett
coefficients satisfying the equations (\ref{e:Bst})
is not assumed a priori; we show in Lemma~\ref{l:Dm} below that
equations (\ref{e:Bst}) express the $d(d+1)/2$ independent
components of $D^{(2)}$ as series not containing any of the $D^{(m)}$,
then the $d(d+1)(d+2)/6$ independent components of $D^{(3)}$ as series
containing only the $D^{(2)}$ and so on.  Lemma~\ref{l:Re} shows that
they are indeed real, and Thm.~\ref{th:conv} shows that the limit
exists.

We define formal power series $f$ and $F$ by 
\be f[{\bf k}]\equiv s[{\bf k}]\Delta T+\i{\bf k}\cdot{\bf a} \label{e:fdef}\ee
\be F[{\bf k}]\equiv\sum_{i=-n}^{n-1}f[{\bf k}]\circ\map^i \label{e:Fdef}\ee
where the dependence on $x$ and on the positive integer $n$
is suppressed in the notation;
the limit $n\rightarrow\infty$ will be taken later.  We have $\la f \ra=0$ and
$\la F \ra=0$ at each order in ${\bf k}$ and for each $n$ as a consequence of
$\la \Delta T \ra=0$ and $\la {\bf a}\ra=0$ above.

We define {\em cumulants} $Q_N[{\bf k}]$ (also formal power series) for 
integers $N\geq2$ as
\begin{equation}\label{e:Qdef}
Q_N[{\bf k}]=\sum_{\{\nu_j\}:\sum_jj\nu_j=N}
(-1)^{\nu-1}\frac{(\nu-1)!\prod_j\la F[{\bf k}]^j \ra^{\nu_j}}
{\prod_j (\nu_j!j!^{\nu_j})}
\end{equation}
with $j$ and $\nu_j$ integers satisfying $j\geq2$ and $\nu_j\geq0$, and
$\nu=\sum_j\nu_j$ is the total number of correlations in the product.
For example
\begin{eqnarray}
Q_2&=&\la F^2 \ra/2\\
Q_3&=&\la F^3 \ra/6\\
Q_4&=&(\la F^4\ra-3\la F^2 \ra^2)/24\\
Q_5&=&(\la F^5 \ra-10\la F^3 \ra \la F^2 \ra)/120\\
Q_6&=&
(\la F^6 \ra-15\la F^4 \ra \la F^2 \ra-10\la F^3 \ra^2+30\la F^2 \ra^3)/720
\label{e:Q6}
\end{eqnarray}
Now $Q_N$ contains exactly $N$ powers of $F$,
and so it contains terms ${\bf k}^m$ only
for $m\geq N$, and we can write it as
\be Q_N[{\bf k}]=\sum_{m=N}^{\infty}\sum_{\alpha_1\ldots\alpha_m}
q_{N,m;\alpha_1\ldots\alpha_m}k_{\alpha_1}\ldots k_{\alpha_m}
\label{e:qdef} \ee
thus defining totally symmetric tensors $q_{N,m}$ for $m\geq N$. 

The Burnett coefficients are found by equating the formal power series on
both sides of
\be s[{\bf k}]=\lim_{n\rightarrow\infty}\frac{1}{2n\la T \ra}\sum_{N=2}^{\infty}Q_N[{\bf k}] \label{e:Bps}\ee
that is,
\be \i^mD^{(m)}_{\alpha_1\ldots\alpha_m}=\lim_{n\rightarrow\infty}
\frac{1}{2n\la T\ra}\sum_{N=2}^{m}q_{N,m;\alpha_1\ldots\alpha_m}
\label{e:Bst}\ee
These equations determine the $D^{(m)}$ explicitly as real tensors,
subject to convergence of the limit, as shown by the following two lemmas.

\begin{lemma}\label{l:Dm}
The right hand side of Eq.~(\ref{e:Bst}) does not contain $D^{(m')}$ such
that $m'\geq m$.
\end{lemma}

Proof: We have $N\geq 2$, and each $Q_N$ contains $N$ powers of $F$, thus
each term has at least 2 powers of $F$.  Each $F$ has at least 1 power of
$\bf k$, and there are $m$ powers of $\bf k$ in total, so each $F$ has
at most $m-1$ powers of $\bf k$.  $D^{(m')}$ appear in $F$ associated
with $m'$ powers of $\bf k$, so $m'\leq m-1$ for any $D^{(m')}$ appearing.

Remark: It is possible there are no factors of $D^{(m')}$ on the right
hand side, in fact the lemma shows that this is true for $m=2$.  The
case $m=2$ can easily be written explicitly; Eq.~(\ref{e:Bst}) becomes
\be \label{e:GK}D^{(2)}_{\alpha\beta}=\lim_{n\rightarrow\infty}
\frac{1}{4n\la T\ra}\sum_{i=-n}^{n-1}\sum_{j=-n}^{n-1}\la a_\alpha^ia_\beta^j \ra \ee
This is a discrete time version of the well-known Green-Kubo formula for
the diffusion tensor (which reduces to a single diffusion coefficient
in the isotropic case $D^{(2)}_{\alpha\beta}=D\delta_{\alpha\beta}$).
An equivalent discrete time equation appears in~\cite{G}, also for $m=4$.

\begin{lemma}\label{l:Re}
Despite the appearance of the imaginary number $\i$ in the above definitions,
the Burnett coefficients are real if they exist.
\end{lemma}

Proof: We note from the definitions that $s[{\bf k}]$, $f[{\bf k}]$ and
$F[{\bf k}]$ have pure imaginary coefficients for odd powers of $\bf k$
and real coefficients for even powers of $\bf k$.  This property is
preserved by addition and multiplication of power series, so it also
holds for the $Q_N[{\bf k}]$.  This implies that the $q_{N,m}$ are
imaginary for odd $m$ and real for even $m$.  The result follows from
Eq.~(\ref{e:Bst}).

Before proceeding with the more technical convergence proof, we note another
important result:

\begin{lemma}\label{l:even}
$D^{(m)}=0$ for $m$ odd.
\end{lemma}

Proof: From the properties of the time reversal operator $\cal T$ given above, 
$\langle F^j\rangle$ has zero contribution from any term with an odd number
of $\bf a$ factors.
The result follows by induction on $m$: assume that $D^{(m')}=0$ for
all odd $m'<m$, then by Lemma~\ref{l:Dm} all terms in $s[{\bf k}]$
contributing to $D^{(m)}$ have even powers of $\bf k$, and from the
oddness of $\bf a$ under time reversal, so also do the $\i{\bf k}\cdot{\bf a}$
terms.  Thus $D^{(m)}$, which is constructed from terms with $m$ powers
of $\bf k$, must be zero for $m$ odd. 

\section{Convergence of the series}\label{s:com}
The averages $\la F^{j} \ra$ appearing in the cumulants contain summations
over $j$ variables with range $-n$ to $n-1$, and could grow as fast as
$O(n^{j})$ in general.  Thus each term, which is a product of such averages
could grow as $O(n^N)$ in general.  For the limit in Eq.~(\ref{e:Bst}) to
exist, we require that the series grows only as $O(n)$.  Although the
growth of each
product of correlations cannot be controlled this well, cancellations
occur in constructing the cumulants.  This is expressed in the following
theorem which, together with Lemmas~\ref{l:Dm} and~\ref{l:Re}, 
implies the existence of the Burnett coefficients:
 
\begin{theorem}\label{th:conv}
$q_{N,m}$ is defined in Eqs.~(\ref{e:Fdef}, \ref{e:Qdef}, \ref{e:qdef})
for integers $N$ and $m$ satisfying $2\leq N\leq m$.  The limit
\be \lim_{n\rightarrow\infty}\frac{1}{n}q_{N,m;\alpha_1\ldots\alpha_m} \ee
exists for all such $N$ and $m$ in the periodic Lorentz gas.
\end{theorem}

The structure of the proof of Thm.~\ref{th:conv} is as follows.  We state the
theorem expressing stretched exponential decay of multiple correlation
functions.  Next, the terms appearing in~(\ref{e:Bst}) are written as a time
ordered sum, so that this theorem can be applied.  Then we show that all the
terms connected by the application of the theorem have coefficients which
sum to zero, so that only the stretched exponential corrections remain.
Finally, a bound of $n$ multiplied by a polynomial is put on the number of
terms at each order of the stretched exponential, so that the series divided by
$n$ converges absolutely.

Thm.~\ref{th:conv} is based on the following result:

\begin{theorem}\label{th:CD}
(Theorem 2 of Ref.~\cite{CD}) Let $i_1\leq\cdots\leq i_k$ and
$1\leq t\leq k-1$.  Then
\begin{equation}
|\la \f{1}\cdots\f{k}\ra-\la \f{1}\cdots\f{t}\ra\la\f{t+1}\cdots\f{k}\ra|
\leq C_k\cdot|i_k-i_1|^2\lambda^{|i_{t+1}-i_t|^{1/2}}
\end{equation}
where $C_k>0$ depends on the functions $f_1,\ldots,f_k$, and $\lambda<1$
is independent of $k$ and $f_1,\ldots,f_k$.
\end{theorem}

The theorem applies to piecewise H\"older continuous functions $f_j$
such that $\la f_j \ra=0$ for all $j$ and uses notation
 $f_j^i\equiv f_j\circ\map^i$.  As noted in
Ref.~\cite{CD}, we expect based on Refs.~\cite{Y,C99} that it should be
possible to prove a stronger bound $\lambda^{|i_{t+1}-i_t|}$, but the
above bound is sufficient for our purposes here.

The $q_{N,m}$ as defined in the previous section are finite sums of terms of
the form (see Eqs.~(\ref{e:Fdef}, \ref{e:Qdef}, \ref{e:qdef}))
\begin{equation}\label{e:series}
\sum_{\{\nu_j\}:\sum_jj\nu_j=N}(-1)^{\nu-1}\frac{(\nu-1)!}
{\prod_j(\nu_j!j!^{\nu_j})}\sum_{i_1\ldots i_N=-n}^{n-1}\la \f{1}\ldots\f{j} \ra
\la \f{j+1}\ldots \ra\ldots\la\ldots \f{N}\ra
\end{equation}
multiplied by constants such as the lower order Burnett coefficients.  The
$f$ here and for the remainder of this section are $T$ or $\bf a$, both
of which satisfy the conditions of Thm.~\ref{th:CD}.  The exact number of
terms of this kind is not important; it depends on $N$ and $m$ but not
$n$ and therefore does not affect convergence of the limit
$n\rightarrow\infty$.

In order to use Thm.~\ref{th:CD} we need to put the times $i_p$ in
numerical order.  The unrestricted sum over all the $i_p$ is replaced
by an ordered sum $i_1\leq i_2\ldots i_N$ over all $N!/S[i]$ permutations
of the $i_p$.  $S[i]$ is a symmetry factor to account for the fact that
some of the $i_p$ may be equal; the exact form is unimportant since it
is a common prefactor, independent of the $\nu_j$.  Not all $N!$ permutations
of the correlations are distinct: it does not matter in which order the
$f_j$ are multiplied within a correlation, or which order correlations of
equal numbers of $f_j$ are multiplied; thus both factorials in
the denominator disappear, leading to
\begin{eqnarray}\nonumber
\sum_{\{\nu_j\}:\sum_jj\nu_j=N}&&(-1)^{\nu-1}(\nu-1)!\left[
\sum_{i_1\leq i_2\ldots i_N}\frac{1}{S[i]}\left\{\la \f{1}\ldots\f{j} \ra
\la \f{j+1}\ldots \ra\ldots\la\ldots \f{N}\ra\right.\right.\\
&&\left.\left.+\mbox{ permutations}\right\}\right]\label{e:sum}
\end{eqnarray}
The ``permutations'' remaining in~(\ref{e:sum}) consist of the remaining
$N!/(\prod_j\nu_j!j!^{\nu_j})-1$ rearrangements of the $i_p$ that are not
equivalent by reordering the product of correlations or the product of
$f$ within a correlation.

As an example, we give the expression for $N=6$:
\begin{eqnarray}\nonumber
&&\sum_{i_1\leq i_2\leq i_3\leq i_4\leq i_5\leq i_6}\frac{1}{S[i]}\left\{
\la\f{1}\f{2}\f{3}\f{4}\f{5}\f{6}\ra\right.\\\nonumber
&&-\left[\la\f{1}\f{2}\ra\la\f{3}\f{4}\f{5}\f{6}\ra
+\la\f{1}\f{3}\ra\la\f{2}\f{4}\f{5}\f{6}\ra
+\la\f{1}\f{4}\ra\la\f{2}\f{3}\f{5}\f{6}\ra\right.\\\nonumber
&&+\la\f{1}\f{5}\ra\la\f{2}\f{3}\f{4}\f{6}\ra
+\la\f{1}\f{6}\ra\la\f{2}\f{3}\f{4}\f{5}\ra
+\la\f{2}\f{3}\ra\la\f{1}\f{4}\f{5}\f{6}\ra\\\nonumber
&&+\la\f{2}\f{4}\ra\la\f{1}\f{3}\f{5}\f{6}\ra
+\la\f{2}\f{5}\ra\la\f{1}\f{3}\f{4}\f{6}\ra
+\la\f{2}\f{6}\ra\la\f{1}\f{3}\f{4}\f{5}\ra\\\nonumber
&&+\la\f{3}\f{4}\ra\la\f{1}\f{2}\f{5}\f{6}\ra
+\la\f{3}\f{5}\ra\la\f{1}\f{2}\f{4}\f{6}\ra
+\la\f{3}\f{6}\ra\la\f{1}\f{2}\f{4}\f{5}\ra\\\nonumber
&&\left.+\la\f{4}\f{5}\ra\la\f{1}\f{2}\f{3}\f{6}\ra
+\la\f{4}\f{6}\ra\la\f{1}\f{2}\f{3}\f{5}\ra
+\la\f{5}\f{6}\ra\la\f{1}\f{2}\f{3}\f{4}\ra\right]\\\nonumber
&&-\left[\la\f{1}\f{2}\f{3}\ra\la\f{4}\f{5}\f{6}\ra
+\la\f{1}\f{2}\f{4}\ra\la\f{3}\f{5}\f{6}\ra
+\la\f{1}\f{2}\f{5}\ra\la\f{3}\f{4}\f{6}\ra\right.\\\nonumber
&&+\la\f{1}\f{2}\f{6}\ra\la\f{3}\f{4}\f{5}\ra
+\la\f{1}\f{3}\f{4}\ra\la\f{2}\f{5}\f{6}\ra
+\la\f{1}\f{3}\f{5}\ra\la\f{2}\f{4}\f{6}\ra\\\nonumber
&&+\la\f{1}\f{3}\f{6}\ra\la\f{2}\f{4}\f{5}\ra
+\la\f{1}\f{4}\f{5}\ra\la\f{2}\f{3}\f{6}\ra
+\la\f{1}\f{4}\f{6}\ra\la\f{2}\f{3}\f{5}\ra\\
&&+\left.\la\f{1}\f{5}\f{6}\ra\la\f{2}\f{3}\f{4}\ra\right]\\\nonumber
&&+2\left[\la\f{1}\f{2}\ra\la\f{3}\f{4}\ra\la\f{5}\f{6}\ra
+\la\f{1}\f{2}\ra\la\f{3}\f{5}\ra\la\f{4}\f{6}\ra
+\la\f{1}\f{2}\ra\la\f{3}\f{6}\ra\la\f{4}\f{5}\ra\right.\\\nonumber
&&+\la\f{1}\f{3}\ra\la\f{2}\f{4}\ra\la\f{5}\f{6}\ra
+\la\f{1}\f{3}\ra\la\f{2}\f{5}\ra\la\f{4}\f{6}\ra
+\la\f{1}\f{3}\ra\la\f{2}\f{6}\ra\la\f{4}\f{5}\ra\\\nonumber
&&+\la\f{1}\f{4}\ra\la\f{2}\f{3}\ra\la\f{5}\f{6}\ra
+\la\f{1}\f{4}\ra\la\f{2}\f{5}\ra\la\f{3}\f{6}\ra
+\la\f{1}\f{4}\ra\la\f{2}\f{6}\ra\la\f{3}\f{5}\ra\\\nonumber
&&+\la\f{1}\f{5}\ra\la\f{2}\f{3}\ra\la\f{4}\f{6}\ra
+\la\f{1}\f{5}\ra\la\f{2}\f{4}\ra\la\f{3}\f{6}\ra
+\la\f{1}\f{5}\ra\la\f{2}\f{6}\ra\la\f{3}\f{4}\ra\\\nonumber
&&\left.\left.+\la\f{1}\f{6}\ra\la\f{2}\f{3}\ra\la\f{4}\f{5}\ra
+\la\f{1}\f{6}\ra\la\f{2}\f{4}\ra\la\f{3}\f{5}\ra
+\la\f{1}\f{6}\ra\la\f{2}\f{5}\ra\la\f{3}\f{4}\ra\right]\right\}
\end{eqnarray}
Here, the four terms correspond to the partitions of $6$ which do not
contain $1$; in the above notation the nonzero $\nu_j$ are
$\{\nu_6=1\}$ with $6!/6!=1$ term;
$\{\nu_2=1,\nu_4=1\}$ with $6!/2!4!=15$ terms;
$\{\nu_3=2\}$ with $6!/2!3!^2=10$
terms; and $\{\nu_2=3\}$ with $6!/3!2!^3=15$ terms; compare with
Eq.~(\ref{e:Q6}). 

Now we apply Thm.~\ref{th:CD} to the largest gap, $i_{t+1}-i_t$.  Any of
the largest gaps will suffice if more than one is largest.  Before
tackling the general case, we see how it works in the $N=6$ example.
Notice that, whatever the value of $t$, the theorem combines all the
above correlations to leave terms (the number of which is a function of $N$)
bounded by $\lambda^{|i_{t+1}-i_t|^{1/2}}$
multiplied by powers of the time differences.  Explicitly, for $t=1$, all
terms cancel individually because $\la f_j \ra=0$.  For $t=2$ the $\la f^6 \ra$
term cancels with one of the $\la f^2 \ra \la f^4 \ra$ terms, six other
$\la f^2 \ra \la f^4 \ra$ terms cancel with three of the $\la f^2 \ra^3$
terms and the remaining terms all split leaving an $\la f \ra$ term.
For $t=3$ the $\la f^6 \ra$ term cancels with one of the $\la f^3 \ra^2$
terms, and all of the others split leaving an $\la f \ra$ term.  $t=4$ is
analogous to $t=2$ and $t=5$ is analogous to $t=1$.

In general we must show that the coefficient $(-1)^{\nu-1}(\nu-1)!$ in
Eq.~(\ref{e:sum}) combined with the numbers of terms of various types leads to
complete cancellation for all values of $N$.  Consider a general term
(ignoring the $S[i]$ which is the same for each term) which is
{\em unaffected} by a split at time $t$.  Each correlation contains 
times $i_p\leq t$ or times $i_p>t$ but not both.  Thus
it can be written schematically as 
\begin{equation}\label{e:ex}
\la \ra \la \ra\ldots\la \ra | \la \ra \la \ra\ldots \la \ra
\end{equation}
where all times $i_p$ to the left of the bar ``$|$'' are less than or equal to
$t$ and all times to the right of the bar are greater than $t$.  Let there be
$A$ correlations to the left and $B$ correlations to the right, so $A+B=\nu$.

This term will cancel (up to stretched exponential corrections) with any term
which is split to the same form, if the sum of the
coefficients (the $(-1)^{\nu-1}(\nu-1)!$) is zero.  The terms that are split
to a given form consist of correlations that are either the same as the
above, or are joined in a pairwise fashion with a correlation on the other
side of the bar.

Again, an example is helpful: When $N=8$, a split at $t=4$ combines the
following terms:
$-6\la\f{1}\f{2}\ra\la\f{3}\f{4}\ra|\la\f{5}\f{6}\ra\la\f{7}\f{8}\ra$ with
$2\la\f{1}\f{2}\f{5}\f{6}\ra\la\f{3}\f{4}\ra\la\f{7}\f{8}\ra$,\\
$2\la\f{1}\f{2}\f{7}\f{8}\ra\la\f{3}\f{4}\ra\la\f{5}\f{6}\ra$,
$2\la\f{1}\f{2}\ra\la\f{3}\f{4}\f{5}\f{6}\ra\la\f{7}\f{8}\ra$,
$2\la\f{1}\f{2}\ra\la\f{3}\f{4}\f{7}\f{8}\ra\la\f{5}\f{6}\ra$,
$-\la\f{1}\f{2}\f{5}\f{6}\ra\la\f{3}\f{4}\f{7}\f{8}\ra$ and
$-\la\f{1}\f{2}\f{7}\f{8}\ra\la\f{3}\f{4}\f{5}\f{6}\ra$.  These all
cancel because $-6+2+2+2+2-1-1=0$.

The term given in Eq.~(\ref{e:ex}) has coefficient $(-1)^{\nu-1}(\nu-1)!$.
There are $AB$ terms with coefficient $(-1)^{\nu-2}(\nu-2)!$ obtained by
combining a single correlation on the left and the right.  There are
$A(A-1)B(B-1)/2!$ terms with coefficient $(-1)^{\nu-3}(\nu-3)!$ obtained
by combining two correlations on the left and the right, and so on until
all ${\rm min}(A,B)$ correlations on the side with the fewest correlations
have been combined.  The total coefficient is thus given by

\begin{equation}\label{e:hyper}
H(A,B)\equiv
\sum_{p=0}^{{\rm min}(A,B)} (-1)^{A+B-p-1}(A+B-p-1)!\frac{A!B!}{(A-p)!(B-p)!p!}
\end{equation}

To show that the coefficients cancel, we therefore need the following lemma:

\begin{lemma}\label{l}
$H(A,B)=0$ for all positive integers $A$ and $B$.
\end{lemma}

{\em Proof:}
The sum is symmetric in $A$ and $B$ so suppose that $A\geq B$ without loss
of generality.  Then
the summand is the product of a constant $(-1)^{A+B-1}A!$, an alternating
binomial of degree $B$, that is, $(-1)^{-p}B!/((B-p)!p!)$ and a polynomial
in $p$ of degree $B-1$, that is, $(A+B-p-1)!/(A-p)!$.  We will use
summation by parts to lower the degree of both until the result is zero. 

We note the summation by parts formula
\begin{equation}\label{e:parts}
\sum_{p=0}^{B}x_py_p=y_0\sum_{p=0}^{B}x_p+\sum_{q=1}^{B}(y_q-y_{q-1})
\sum_{p=q}^{B}x_p
\end{equation}
which can be demonstrated by collecting terms on the right hand side.  Now
substituting $x_p=(-1)^{-p}B!/((B-p)!p!$ and $y_p=(A+B-p-1)!/(A-p)!$
we can show by induction on $q$ from $B$ downwards that
\begin{equation}
\sum_{p=q}^{B}x_p=\left\{
\begin{array}{cc}(-1)^{-q}\frac{(B-1)!}{(B-q)!(q-1)!}&q>0\\
0&q=0\end{array}\right.
\end{equation}
hence the first term on the right hand side of Eq.~(\ref{e:parts}) vanishes.
We can also simplify
\begin{equation}
y_q-y_{q-1}=(1-B)\frac{(A+B-q-1)!}{(A-q+1)!}
\end{equation}
so Eq.~(\ref{e:hyper}) now reads
\begin{equation}
H(A,B)=(-1)^{A+B-1}A!(1-B)\sum_{q=1}^B\frac{(-1)^{-q}(B-1)!}{(B-p)!(p+1)!}
\frac{(A+B-q-1)!}{(A-q+1)!} 
\end{equation}
Shifting the summation index by one we find
\begin{equation}
H(A,B)=(1-B)H(A,B-1)
\end{equation}
The proof of Lemma~\ref{l} follows by noting that $H(A,1)=0$.

We now conclude the proof of Thm.~\ref{th:conv}.  Recall that the
series~(\ref{e:series}) have been rewritten in the form~(\ref{e:sum}).
Thm.~\ref{th:CD} is applied to (one of) the largest gap(s) 
$\Delta i_{\rm max}\equiv i_{t+1}-i_t$,
partitioning the terms into subsets which split into a particular form
(\ref{e:ex}).  Lemma~\ref{l} shows that the coefficients of all terms
in a subset conspire to cancel, so that each subset is bounded by the
error term in theorem~\ref{th:CD}, that is
$\lambda^{|\Delta i_{\rm max}|^{1/2}}$
multiplied by a polynomial in the time differences.

Finally we estimate the number of terms with each value of
$\Delta i_{\rm max}$.  The first time $i_1$ varies freely from $-n$ to
$n-1$, having a total of $2n$ values.  One of the time differences
is equal to $\Delta i_{\rm max}$, and the other $k-2$ time differences
can range from $0$ to $\Delta i_{\rm max}$, so the total number of terms
with a given $\Delta i_{\rm max}$ is less than
$2n(k-1)\Delta i_{\rm max}^{(k-2)}$, in particular a polynomial in
$\Delta i_{\rm max}$ multiplied by $n$.  Thus the series divided by $n$
appearing in Thm.~\ref{th:conv} is bounded by a product of polynomial
factors and the decaying stretched exponential, and hence converges absolutely.
This concludes the proof of Thm.~\ref{th:conv} and the proof of existence
of Burnett coefficients.

\section{Physical motivation and remarks}
This section makes the connection between the Burnett coefficients defined
in the previous sections and equations found in the physics literature.
The latter equations
are phenomenological and have not been shown rigorously in a limiting
fashion from the Lorentz gas, and a few nonrigorous
limit interchanges are made to connect them with the expressions defined
in the previous sections.  First we consider the dispersion relation,
the equation for the Burnett coefficients, and finally the whether the
dispersion relation can be used to define an analytic function.

The dispersion relation~(\ref{e:disp})
with $\bf k$ interpreted as a real vector represents the solution
of a generalised diffusion equation proposed by Burnett~\cite{B}
containing higher derivative terms that become important on small scales,
\begin{equation}\label{e:dgen}
\partial_t\rho=\sum_{m=2}^{\infty}\sum_{\alpha_1\ldots\alpha_m}
D^{(m)}_{\alpha_1\ldots\alpha_m}
\partial_{\alpha_1}\ldots\partial_{\alpha_m}\rho
\end{equation}
assuming a solution of the form
\begin{equation}\label{e:wave}
\rho({\bf r},t)\sim\exp(s({\bf k})t+\i {\bf k\cdot r})
\end{equation}
Here, $\partial_\alpha\equiv \partial/{\partial r_\alpha}$.
Nonlinear terms such as powers of $\partial_\alpha\rho$ are excluded on physical
grounds since $\rho$ is a projection onto real space (${\bf r}\in\IR^d$) of a
phase space density satisfying a linear evolution equation.
The {\em phase space}
is a subset of $\IR^{2dM}$ corresponding to the possible positions and
velocities of $M\gg 1$ particles.  The dispersion
relation is a more robust formulation than the generalized diffusion
equation~(\ref{e:dgen}) since the former may be supplemented by
nonanalytic functions
of $\bf k$ to account for situations (other than the periodic Lorentz
gas) in which some of the Burnett coefficients do not exist.

Chapter 7 of Ref.~\cite{G} obtains the dispersion relation from the
microscopic dynamics using the equation (7.91 in this reference):
\begin{equation}
1=\lim_{n\rightarrow\infty}\langle\prod_{i=-n}^{n-1}\exp[
-s({\bf k})T(\map^ix)-\i{\bf k}\cdot{\bf a}(\map^ix)]\rangle
\end{equation}
We write $T=\langle T\rangle+\Delta T$ as in previous sections,
take out the constant factor of $\langle T\rangle$, and take the
logarithm to find
\begin{equation}
s({\bf k})=\lim_{n\rightarrow\infty}\frac{1}{2n\langle T \rangle}
\ln\langle\exp[F({\bf k})]\rangle
\end{equation}
where $F$ is defined (as a power series) in Eq.~(\ref{e:Fdef}).  Now
the exponential and the logarithm are expanded in power series and
the resulting terms containing $N$ powers of $F$ are collected to
become the cumulants $Q_N$ defined in Eq.~(\ref{e:Qdef}).  The cumulant
form of the expansion is possibly more robust than the above equations
due to the cancellations among the terms that combine to construct
each cumulant.

Since it is desirable from a physical point of view to interpret $\bf k$
as a real variable, we conclude with the following conjecture:

\begin{conjecture}
The series~(\ref{e:disp}) converges when ${\bf k}\in{\cal D}\subset\IR^d$
for some nontrivial domain $\cal D$,
and so defines a function $s(k)$ in this domain.
\end{conjecture} 

Note that $s(k)$ (if it exists) is a real function as a consequence of
Lemma~\ref{l:even}, and physically is expected to be negative except at
the origin (otherwise the density $\rho$ would grow exponentially with time);
this puts further constraints on the Burnett coefficients.

Unfortunately the proof given in the previous section
contains many undetermined functions of $\bf k$, and the Burnett
coefficients are defined by a complicated recursive relation~(\ref{e:Bst}),
so a proof is unlikely using the techniques of this paper.

There are two results that make such a result plausible.  The first is
that in the Boltzmann limit of a hard sphere gas, that is, a gas with
many moving particles at low density and with recollisions ignored,
the expansion in $\bf k$ (in this context called the linearized
Chapman-Enskog expansion) converges~\cite{McL}.  Of course, the hard
sphere collisions are similar to that of the Lorentz gas, but recollisions
cannot be ignored in general. 
 
The second result is exact, but for a highly simplified (piecewise linear)
system.  We consider the map $\map:\IR\to\IR$ given by
\begin{equation}
\map(x)=\frac{3}{2}-2x+3[x]
\end{equation}
where $[x]$ is the greatest integer less than or equal to $x$.
The dynamics defined by $\map$ is equivalent to a random walk
where the particle moves with equal probability
from one interval $I_n\equiv(n-1/2,n+1/2)$ to the left, $I_{n-1}$ or to
the right, $I_{n+1}$.  The dispersion relation $s(k)$ follows directly
from the above phenomenological solution~(\ref{e:wave}), 
\begin{equation}
\rho(n,t)=\exp(st+ikn)
\end{equation}
After one iteration,
\begin{eqnarray}
\rho(n,1)&=&\frac{1}{2}[\exp(\i{}k(n-1))+\exp(\i{}k(n+1))]\\
&=&\cos{k}\exp(\i{}kn)
\end{eqnarray}
leading to
\begin{equation}
s(k)=\ln\cos{k}
\end{equation}
which has a power series around $k=0$ with a radius of convergence equal
to $\pi/2$.


\begin{thebibliography}{99}
\bibitem{BSC}L. A. Bunimovich, Ya. G. Sinai and N. I. Chernov.
Statistical properties of two-dimensional hyperbolic billiards
{\em Russ. Math. Surv.} {\bf 46} (1991) 47-106.
\bibitem{B}D. Burnett.  The distribution of molecular velocities and the
mean motion in a non-uniform gas,
{\em Proc. London Math. Soc.} {\bf 40} (1935) 382-435.
\bibitem{C94}N. I. Chernov. Statistical properties of the periodic Lorentz gas.
Multidimensional case. {\em J. Stat. Phys.} {\bf 74} (1994) 11-53.
\bibitem{C99}N. I. Chernov.  Decay of correlations and dispersing billiards.
{\em J. Stat. Phys.} {\bf 94} (1999) 513-556.
\bibitem{CD}N. I. Chernov and C. P. Dettmann.  The existence of Burnett
coefficients in the periodic Lorentz gas. {\em Physica A} {\bf 279} (2000)
37-44.
\bibitem{DC}C. P. Dettmann and E. G. D. Cohen.  Microscopic chaos
and diffusion. {\em J. Stat. Phys.} {\bf 101} (2000) 775-817.
\bibitem{G}P. Gaspard {\em Chaos, scattering and statistical mechanics}
Cambridge University: Cambridge, 1999.
\bibitem{L}H. A. Lorentz.  The motion of electrons in metallic bodies.
{\em Proc. Amst. Acad.} {\bf 7} (1905) 438-453.
\bibitem{McL}J. A. MacLennan.  {\em Introduction to non-equilibrium statistical
mechanics.} Prentice-Hall: London, 1989 pp144-145.
\bibitem{S}D. Szasz (ed.) {\em Hard ball systems and the Lorentz gas}
Springer: Heidelberg, 2000.
\bibitem{vB}H. van Beijeren.  Transport properties of stochastic Lorentz
models. {\em Rev. Mod. Phys.} {\bf 54} (1982) 195-234.
\bibitem{Y}L.-S. Young.  Statistical properties of dynamical systems with
some hyperbolicity. {\em Annals of Math.} {\bf 147} (1998) 585-650.
\end{thebibliography}
\end{document}